\documentclass[a4paper,twocolumn,showpacs,preprintnumbers,amsmath,amssymb,nofootinbib]{revtex4}
\input psfig.sty 
\usepackage{graphicx}

\begin{document}

\title{Current constraints on the epoch of cosmic acceleration}

\author{B. Santos$^1$}

\author{J. C. Carvalho$^2$}

\author{J. S. Alcaniz$^1$}

\address{$^1$Departamento de Astronomia, Observat\'orio Nacional, 20921-400, Rio de Janeiro - RJ, Brasil}

\address{$^2$Departamento de F\'isica, Universidade Federal do Rio Grande do Norte, 59072-970, Natal - RN, Brasil}

\date{\today}

\begin{abstract}
The cosmographic expansion history of the universe is investigated by using the 557 type Ia supernovae from the Union2 Compilation set along with the current estimates involving the product of the CMB acoustic scale $\ell_{A}$ and the BAO peak at two different redshifts. Using a well-behaved parameterization for the deceleration parameter, $q(z) = q_0 + q_1z/(1 + z)$, we estimate the accelerating redshift $z_{acc}=-q_0/(q_0 + q_1)$ (at which the universe switches from deceleration to acceleration) and investigate the influence of a non-vanishing spatial curvature on  these estimates. We also use the asymptotic value of $q(z)$ at high-$z$ to place more restrictive bounds on the model parameters $q_0$ and $q_1$, which results in a more precise determination of the epoch of cosmic acceleration.

\end{abstract}


\maketitle


\emph{Introduction}--Although a late-time cosmic acceleration is becoming observationally well established (at $> 12\sigma$ from some independent analyses~\cite{cm}), the physical mechanism behind this phenomenon constitutes a completely open question nowadays in Cosmology.  In principle, this accelerating expansion may be the result of unknown physical processes involving either the existence of new fields in high energy physics, the so-called dark energy in the context of general relativistic models~\cite{review}, or the infrared modification of gravity one should expect, e.g., from extra dimensional physics, which would lead to a modification of the effective Friedmann equation at late times~\cite{bw,sahni}. Another possibility of modification of gravity is to add terms proportional to powers of the Ricci scalar $R$ to the Einstein-Hilbert Lagrangian, the so-called $f(R)$ gravity models~(see, e.g., \cite{fr} and references therein).

Given the current state of uncertainty that remains over which mechanism may provide a complete description for cosmic acceleration, a better understanding of some characteristics of this phenomenon can certainly  help us distinguish among many alternative models of universe. This is the case, for instance, of the duration of the accelerating phase, which depends crucially on the physical  mechanism of acceleration. In principle, it may distinguish between thawing and freezing potentials~\cite{prl} or between some classes of braneworld models and the standard $\Lambda$CDM scenario~\cite{sahni}, and may also test the compatibility between classes of dark energy models and theoretical constraints from  String/M theory\footnote{As discussed in Ref.~\cite{string}, an eternally accelerating universe seems not to be in agreement with String/M-theory predictions, since it is endowed with a cosmological event horizon which prevents the construction of a conventional S-matrix describing particle interactions. This is the case of freezing scenarios as well as of our current standard $\Lambda$CDM model.}. 

Another important aspect in this discussion concerns the epoch of the transition from an initially decelerated to an accelerating phase. A purely kinematic record of the expansion history of the universe, without regard to its cause or the validity of any particular metric theory of gravity, can provide answers to basic questions concerning the acceleration history and model-independent constraints with which one may test cosmological models. As an example, we note that one of the  features of evolving dark energy is that the redshift at which the universe switches from deceleration to acceleration, $z_{acc}$, is remarkably different from that in $\Lambda$CDM scenario with the same amount of dark energy today, a fact that is manifested in the Cosmic Microwave Background (CMB) data as a modified integrated Sachs-Wolfe effect~\cite{bassett}. 

Our goal in this \emph{Brief Report} is to derive updated cosmographic bounds on the epoch of the cosmic acceleration, without making any assumption on the energy content of the Universe (for similar analyses using either different parameterizations and expansion formulas for the deceleration parameter $q(z)$ or different data sets, see~\cite{cm,bassett,qz}). Our approach consists on a parametric approximation of the deceleration parameter along the cosmic evolution, given by $q(z)= q_o + q_1z/(1+z)$~\cite{cm}.  In order to ensure a period of structure formation during the matter-dominated era, we also make use of the asymptotic value of $q(z)$ at high redshift to constrain the model parameters $q_0$ and $q_1$. The statistical analysis is performed using the most recent SNe Ia observational data, the so-called Union2 sample of 557 events~\cite{union2}, together with the recent estimates involving the product of the CMB acoustic scale $\ell_{A}$ and the baryonic acoustic oscillation (BAO) peak~\cite{sollerman}. In what follows, we outline the main assumptions for our analysis and discuss our main results.


\emph{Basic equations and Analysis}--We start our discussion by assuming the Friedmann-Lema\^itre-Robertson-Walker (FLRW) line element. In such a background, the comoving distance to a given redshift $z$ can be written as 
\begin{equation}
r(z) = \frac{cH_0^{-1}}{\sqrt{|\Omega_{k0}|}} S_k\left[\sqrt{|\Omega_{k0}|} \int_{0}^{z} {\cal{I}}(u) du \right] \;,
\end{equation}
where ${\cal{I}}(u) = \exp{\left\{ -\int_{0}^{u} [1+q(u)]d\ln({1+u}) \right\}}$ and the function $S_k$ is defined by one of the following
forms: $S_k(r) = \sinh(r)$, $r$, and $\sin(r)$ for open, flat and closed geometries, respectively. In the above equation, we have used the definition of the deceleration parameter $q = -{\ddot{a}a}/{\dot{a}^2}$ or, equivalently, 
\begin{equation}\label{qh}
q(z) = \frac{1}{2}\frac{d\ln{H^2}}{d\ln{(1+z)}} -1\;,
\end{equation}
where $a(t) = (1+z)^{-1}$ stands for the cosmological scale factor, $H = \dot{a}/a$ is the Hubble parameter and a dot denotes derivatives with respect to time. 

In order to proceed further, we follow~\cite{cm} and adopt in our analysis a two-parameter expression for the deceleration parameter 
\begin{equation} \label{qz} 
q(z) = q_0 + q_1 \frac{z}{1+z} \;,
\end{equation}
where $q_0$ is the current value of the deceleration parameter and in the infinite past $q(\infty) = q_0 + q_1$. Clearly, this parametric form for $q(z)$ was inspired from one of the most popular  parametrization of the dark energy equation of state~\cite{cpl} and, although very simple, seems to be flexible enough to mimic the $q(z)$ behavior of wide class of accelerating models~\cite{sollerman}. From Eq. (\ref{qh}), we can also calculate the Hubble parameter for the above parametrization, i.e.,
\begin{equation} \label{hz}
H(z) = H_0(1+z)^{1+q_0+q_1} \exp\left(-q_1{z\over 1+z}\right)\;.
\end{equation}
The accelerating redshift ($q(z_{acc}) = 0$) for parameterization (\ref{qz}) is given by 
\begin{equation}
	\label{eq:model.zt}
	z_{acc} = -\frac{q_0}{q_0 + q_1} \,.
\end{equation}


In order to derive constraints on the epoch of cosmic acceleration, we use in our analyses the most recent SNe Ia compilation available, namely, the Union2 sample of Ref.~\cite{union2}. This compilation 
is an update of the original Union sample that comprises 557 data points in the redshift range $0.015 < z < 1.4$ including recent large samples from other surveys and using SALT2 for SN Ia light-curve fitting.

The predicted distance modulus for a supernova at redshift $z$, given a set of
parameters $\mathbf{s}$, is
$\mu_p(z|\mathbf{s}) = m - M = 5\mbox{log} d_L + 25$,
where $m$ and $M$ are, respectively, the apparent and absolute magnitudes and $d_L$ stands for the luminosity distance,
$d_L = (1 + z)r(z)$ (in units of megaparsecs).
In our analysis, we estimated the best fit to the set of parameters $\mathbf{s} \equiv (H_0, q_0, q_1)$ by using a $\chi^{2}$
statistics
$\chi_{\rm{SNe}}^{2} = \sum_{i=1}^{557}{{\left[\mu_p^{i}(z|\mathbf{s}) -\mu_o^{i}(z)\right]^{2}}/{\sigma_i^{2}}}$,
where $\mu_p^{i}(z|\mathbf{s})$ is given above, $\mu_o^{i}(z)$ is the extinction corrected distance modulus for a given SNe Ia at $z_i$, and $\sigma_i$ is the uncertainty in the individual distance moduli. The Hubble parameter $H_o$ is considered a nuisance parameter so that we marginalize over it by using the analytical method of Ref. \cite{wang}.

Along with the SNe Ia data, and to impose more restrictive bounds on the parameters $q_0$ and $q_1$, we use the two major inputs involving acoustic oscillations that come from the CMB and baryon oscillations data. We follow Ref.~\cite{sollerman} and use constraints derived from the product of the CMB acoustic scale $\ell_{A} = \pi d_A (z_*)/r_s(z_*)$ and the measurement of the ratio of the sound horizon scale at the drag epoch to the BAO dilation scale, $r_s(z_d )/D_V(z_{\rm{BAO}})$. In the above expressions, $d_A (z_*)$ is the comoving angular-diameter distance to recombination $z_* = 1089$ and $r_s(z_*)$ is the comoving sound horizon at photon decoupling given by
	$r_s(z_*) = \int_{z_*}^{\infty} \frac{c_s}{H(z)} dz$,
which depends upon the speed of sound before recombination, $c_s$. $z_d \simeq 1020$ is the redshift of the drag epoch (at which the acoustic oscillations are frozen in) and the so-called dilation scale, $D_V$, is
given by 
$D_V(z) = [czr^{2}(z)/H(z)]^{1/3}$.
By combining the ratio $r_s (z_d = 1020)/r_s (z_*=1090) = 1.044 \pm 0.019$~\cite{komatsu10} with the measurements of $r_s(z_d )/D_V(z_{\rm{BAO}})$ at $z_{\rm{BAO}} = 0.20$ and 0.35 from Ref.~\cite{percival}, Sollerman {\it et al.}~\cite{sollerman} found 
$$
f_{0.20} = d_A (z_*)/D_V (0.2) = 17.55 \pm0.65 
$$
$$
f_{0.35} = d_A (z_*)/D_V (0.35) = 10.10 \pm 0.38\;.
$$
In order to derive the constraints on acceleration redshift $z_{acc}$ discussed in the next section we minimize the function $\chi^2_{\rm{T}} = \chi^2_{\rm{SNe}} + \chi^2_{\rm{CMB/BAO}}$, where $\chi_{\rm{CMB/BAO}}^{2} = \left[f_{0.2}(z|\mathbf{s}) - f_{0.2}\right]^2/\sigma_{0.2}^2 + \left[f_{0.35}(z|\mathbf{s}) - f_{0.35}\right]^2/\sigma_{0.35}^2$. 


\emph{Results}--The results of our analysis are shown in Figs. 1 and 2 and in Table 1. Figure 1a shows 68.3\% and 95.4\% confidence regions in the ($q_0$, $q_1$) plane for parameterization (\ref{qz}) arising from Union2 SNe Ia sample (blue ellipses) and measurements of CMB/BAO ratio (green bands) when vanishing spatial curvature ($\Omega_{k0} = 0$) is assumed. Note that, although not very restrictive when applied separately, these SNe Ia and CMB/BAO contours constrain different regions in the parametric space, showing the complementarity of these data sets, and resulting in very tight bounds on the ($q_0$, $q_1$) space when used together. Quantitatively, if we use only SNe Ia data we obtain $z_{acc} = 0.75 \pm 0.35$ (2$\sigma$) whereas the CMB/BAO measurements provide $z_{acc} = 1.20 \pm 0.10$ (2$\sigma$). From the SNe Ia + CMB/BAO analysis, we find $z_{acc} = 0.87 \pm 0.10$ (2$\sigma$) with reduced $\chi^2_{\nu} \simeq 0.97$, which is roughly consistent with the $\Lambda$CDM prediction with $\Omega_m = 0.27 \pm 0.04$~\cite{komatsu10}, i.e., $z_{acc} \in [0.64,0.88]$.  The constraining power of the joint SNe Ia + CMB/BAO analysis is clearly shown by the $2\sigma$ black contours. In Fig. 1b, we show the evolution of the deceleration parameter $q(z)$ for the resulting $2\sigma$ intervals of $q_0$ and $q_1$ obtained using the data sets considered above. We note that the general behavior of $q(z)$ is very similar to the one predicted by several classes of accelerating scenarios and reflects the flexibility of parameterization (\ref{qz}) discussed earlier. For comparison, the $\Lambda$CDM curve  with $\Omega_m = 0.27$ is also shown (thick line).


For most of the viable cosmological scenarios, we expect the Universe to be matter-dominated at early times (i.e., after the radiation dominance), which implies $q=1/2$. Thus, in order to ensure this past dark matter-dominated epoch (and the standard $a(t) \propto t^{2/3}$ law), whose existence is fundamental for the structure formation process to take place, we assume the constraint $q(z>>1) = 1/2$. In this case, $q_0 + q_1 = 1/2$ and Eq.~(\ref{qz}) becomes
\begin{equation}
	\label{eq:stat.paramet.cond}
	q(z) = \left(q_0 + \frac{z}{2}\right) \frac{1}{1+z}\;,
\end{equation}
which has the advantage of reducing our analysis to one-parameter fitting. From this equation, the acceleration redshift is now simply given by
$z_{acc} = -2q_0$. 
Note also that, for $z >> 1$, Eq.~(\ref{hz}) reduces to $H \propto (1+z)^{3/2}$, as expected for a decelerating matter-dominated universe.

Figure 2a shows the behavior of the deceleration parameter as a function of $z$ when the above constraint is taken into account. Note that, differently from the results shown in Panel 1b, the CMB/BAO estimate on $z_{acc}$ is now moved to a much lower value, i.e., $z_{acc} = 0.32 \pm 0.20$ (2$\sigma$) whereas the Union2 sample alone gives $z_{acc} = 1.16 \pm 0.18$ (2$\sigma$). The joint analysis involving these two data sets plus the above constraint on $q(z)$ at high-$z$ limit provides $z_{acc} = 0.71 \pm 0.12$ (2$\sigma$). This is in good agreement with the concordance $\Lambda$CDM model prediction with $\Omega_m = 0.27$ ($z_{acc} = 0.75$) as well as with a $w$CDM model\footnote{The accelerating redshift in $w$CDM models with a constant EoS parameter $w$ can be obtained from the expression:
$\Omega_m(1+z_{acc})^3 + (1+3w)(1-\Omega_m)(1+z_{acc})^{3(1+w)} = 0$, 
which reduces to the $\Lambda$CDM prediction  $z_{acc} = (2\Omega_{\Lambda}/\Omega_m)^{1/3} - 1$ when $w = -1$.} with the same amount of non-relativistic matter today and equation-of-state (EoS) parameter ranging in the interval $-1.76 \leq w \leq -0.63$.
\begin{figure}[t]
\centerline{\psfig{figure=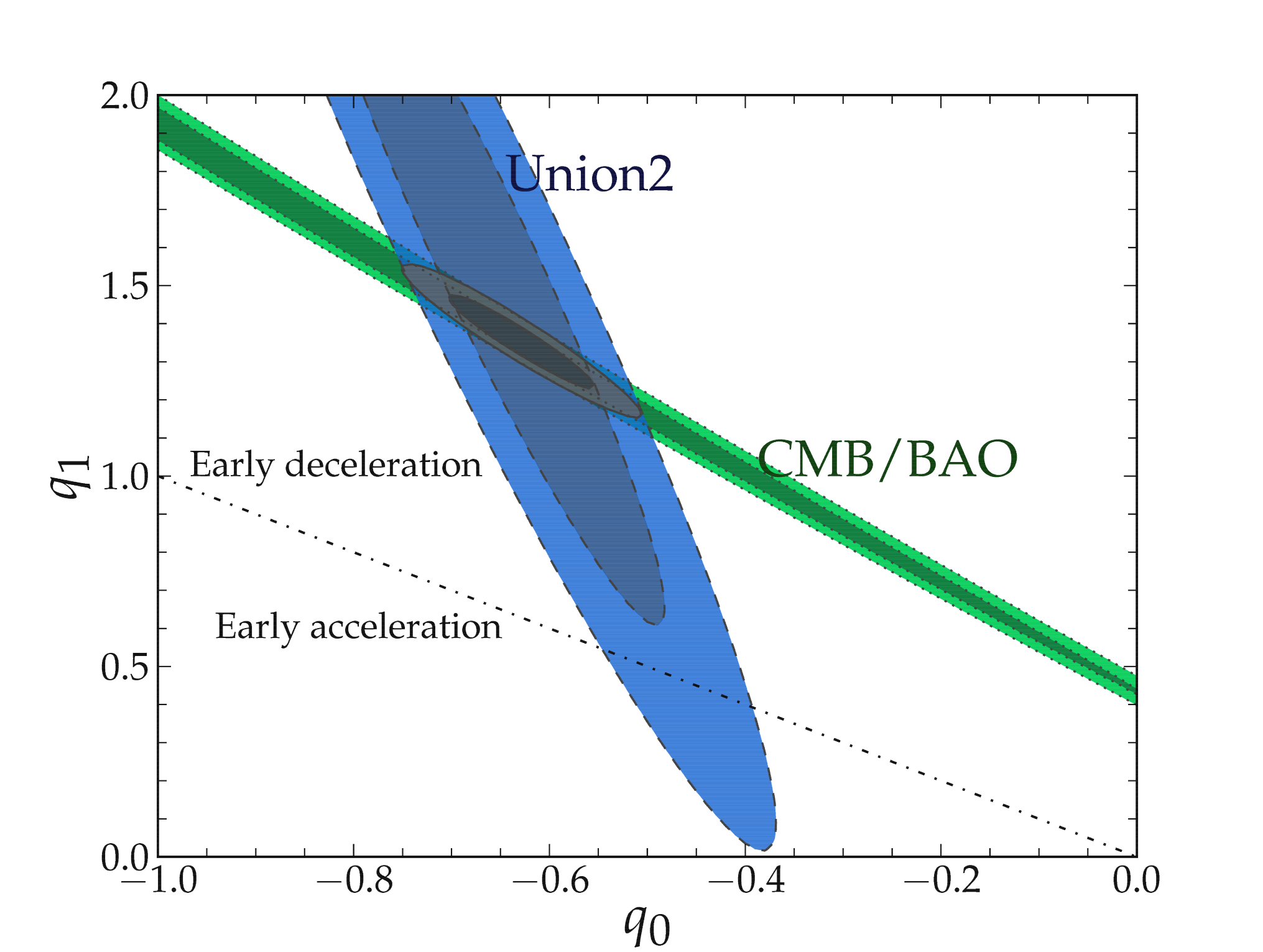,width=1.7truein,height=2.3truein}
\psfig{figure=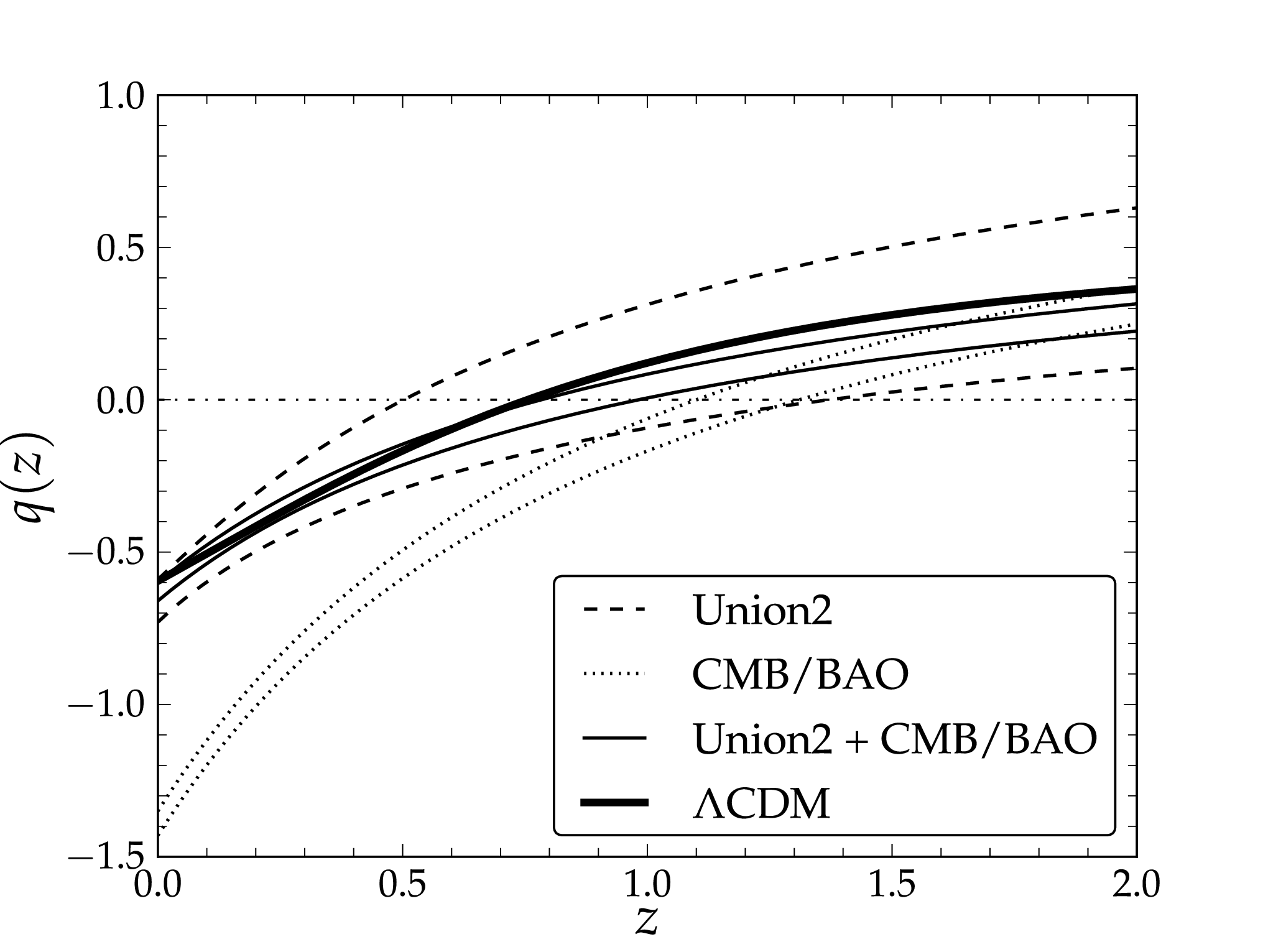,width=1.7truein,height=2.3truein} 
\hskip 0.1in} 
\caption{{\bf{a)}} Confidence regions in the  ($q_0$, $q_1$) plane for parameterization (\ref{qz}) arising from Union2 SNe Ia sample (blue ellipses) and measurements of CMB/BAO ratio (green bands) for $\Omega_{k0} = 0$. Black regions correspond to the resulting bounds from the joint analysis. Contours are drawn for $\Delta \chi^2 = 2.30$ and $\Delta \chi^2 = 6.17$. {\bf{b)}} Evolution of the deceleration parameter $q(z)$ for the resulting $2\sigma$ intervals of $q_0$ and $q_1$ obtained using the data sets discussed in the text.}
\end{figure}

\begin{figure}[t]
\centerline{\psfig{figure=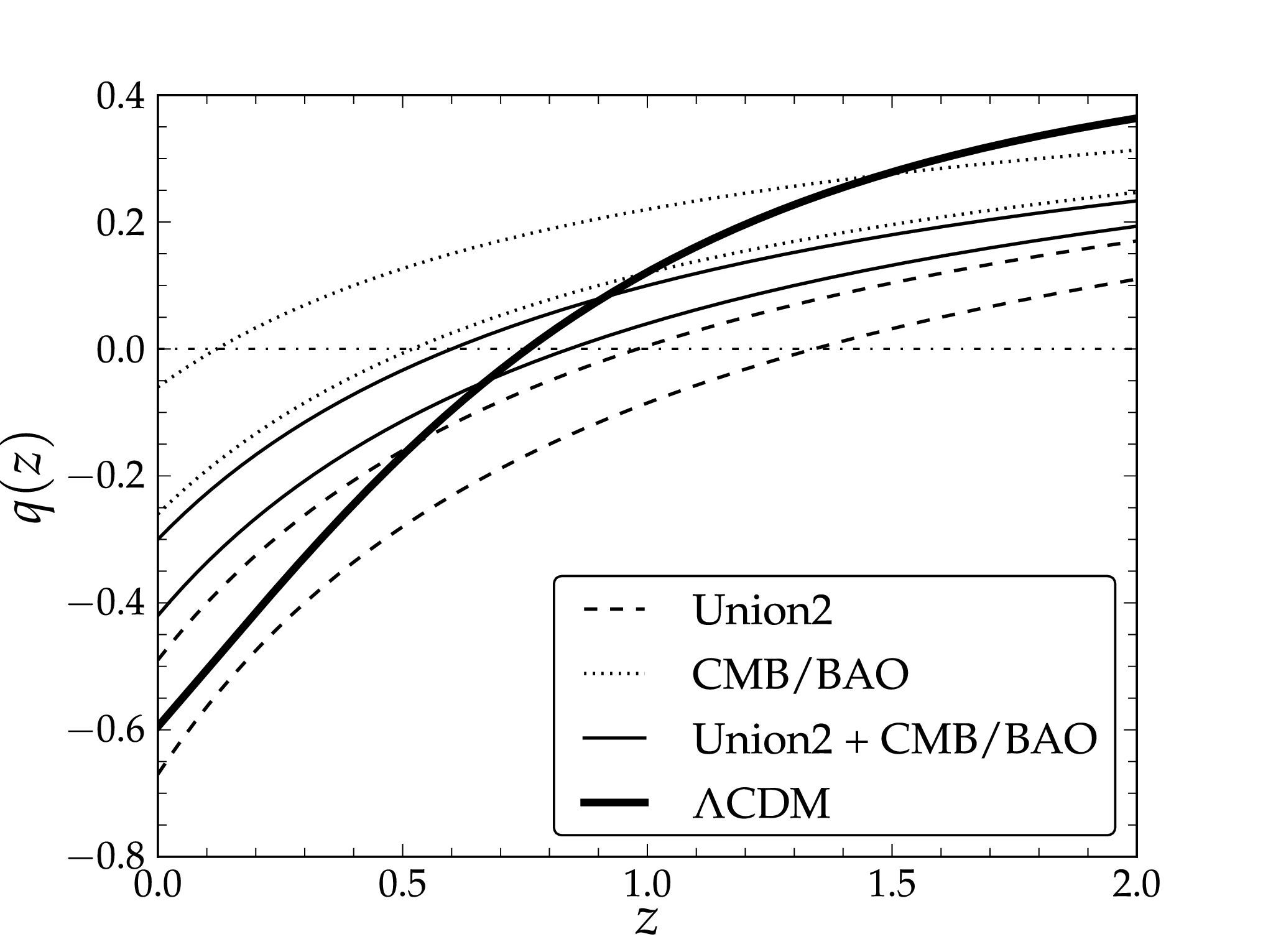,width=1.7truein,height=2.3truein}
\psfig{figure=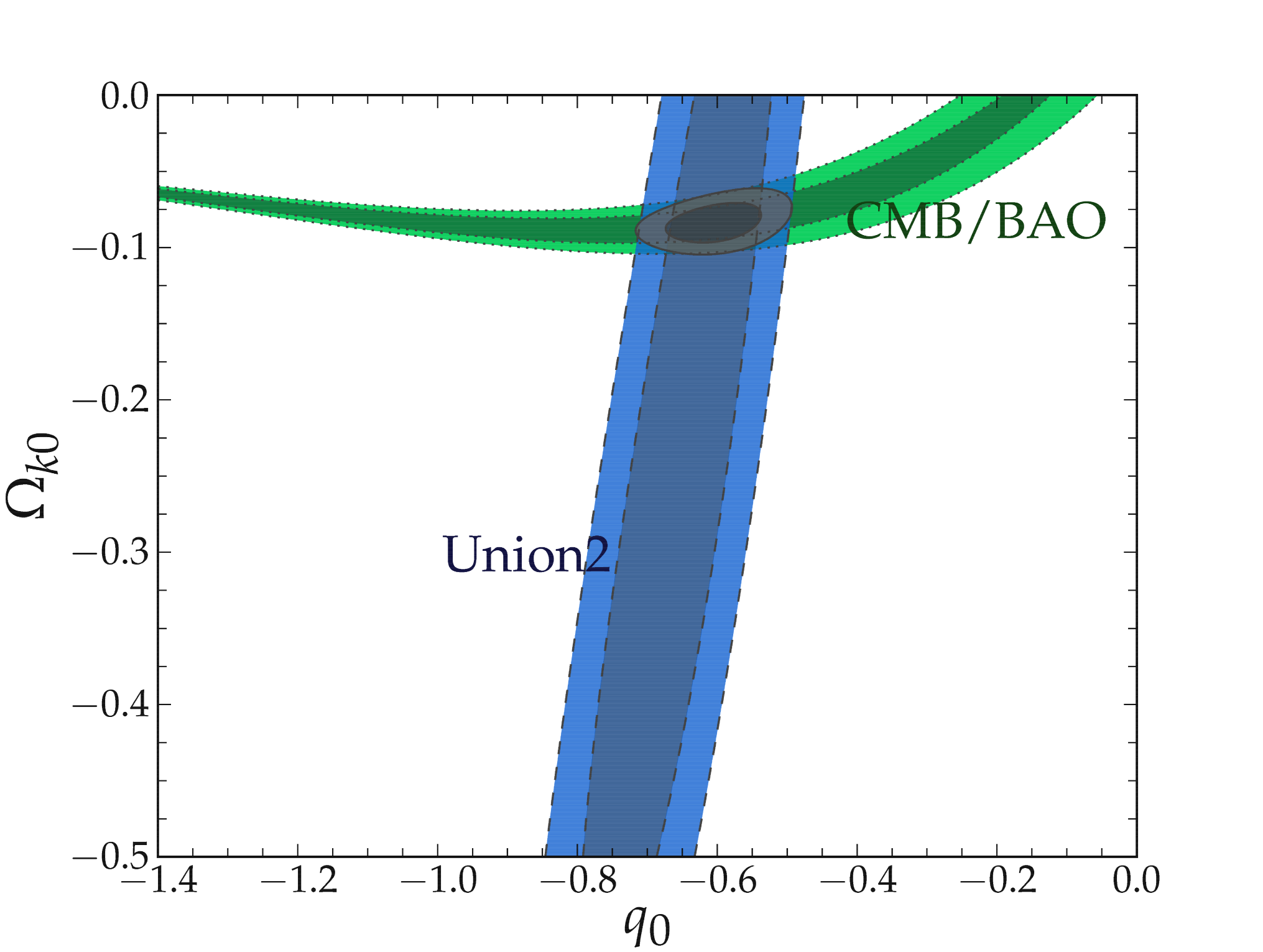,width=1.7truein,height=2.3truein} 
\hskip 0.1in} 
\caption{{\bf{a)}} The same as in Fig. 1b when the high-$z$ limit on the deceleration parameter $q(z>>1) = 1/2$ is considered. {\bf{b)}} Confidence regions in the  ($q_0$, $\Omega_{k0}$) space from SNe Ia + CMB/BAO +  high-$z$ limit on $q(z)$. Note that a spatially closed and current accelerating universe is favored by the joint analysis (black contours). All the contours are drawn for $\Delta \chi^2 = 2.30$ and $\Delta \chi^2 = 6.17$.}
\end{figure}

Further, by considering the  above high-$z$ limit of $q(z)$, we show in Fig. 2b the effect of a non-vanishing spatial curvature on the $z_{acc}$ estimates. From this analysis, we see that SNe Ia and CMB/BAO data provide almost orthogonal bounds on the ($q_0,\Omega_{k0}$) plane and, therefore, very tight constraints on the current values of the deceleration and curvature parameters. In particular, we found $q_0 = -0.61\pm0.08$ and $\Omega_{k0} =-0.08 \pm 0.02$ at $2\sigma$ level, which in turn favors a spatially closed and current accelerating universe, with $z_{acc} = 1.21 \pm 0.17$ ($2\sigma$).  

For the sake of completeness, we also derive  the accelerating redshift by assuming  the current 1$\sigma$ bounds on the curvature parameter from WMAP7 results, $\Omega_{k0} = -0.0023_{-0.0056}^{+0.0054}$~\cite{komatsu10}. As can be seen from Table I, the effect of curvature for this realistic interval of $\Omega_{k0}$ is small, being $\lesssim 8\%$  for the two extreme cases of open ($\Omega_{k0} = 0.0031$) and closed ($\Omega_{k0} = -0.0079$) geometries. The main quantitative results of this paper are summarized in Table 1.

\begin{table*}[]  
\begin{center}  
\caption{Current limits on $z_{acc}$ from SNe Ia and CMB/BAO data.}  
\begin{tabular}{rrllrl}  
\hline  \hline \\
\multicolumn{1}{c}{Test}&
\multicolumn{1}{c}{$q_0$}& 
\multicolumn{1}{c}{$q_1$}& 
\multicolumn{1}{c}{$z_{acc}$}& 
\multicolumn{1}{c}{$\chi^2_{min}$}\\   
\hline 
& & \quad \quad \quad \quad {$\Omega_{k0} = -0.0079$}& & \\
\hline 
CMB/BAO.....& $-1.36\pm0.02(1\sigma)\pm0.04(2\sigma)$	& $2.48\pm0.03(1\sigma)\pm0.07(2\sigma)$	& $1.21\pm0.05(1\sigma)\pm0.11(2\sigma)$	& $1.18 \times 10^{-4}$ \\
Union2....& $-0.66\pm0.03(1\sigma)\pm0.07(2\sigma)$	& $1.53\pm0.19(1\sigma)\pm0.38(2\sigma)$	& $0.76\pm0.18(1\sigma)\pm0.36(2\sigma)$	& 543.22 \\
Union2 + CMB/BAO....& $-0.63\pm0.01(1\sigma)\pm0.03(2\sigma)$	& $1.34\pm0.02(1\sigma)\pm0.05(2\sigma)$	& $0.89\pm0.04(1\sigma)\pm0.10(2\sigma)$	& 543.97 \\
\hline
& & \quad \quad  \quad \quad{$\Omega_{k0} = 0.0031$}& &\\
\hline
CMB/BAO.....& $-1.36\pm0.02(1\sigma)\pm0.04(2\sigma)$& $2.51\pm0.03(1\sigma)\pm0.07(2\sigma)$	& $1.18\pm0.05(1\sigma)\pm0.10(2\sigma)$	& $5.68 \times 10^{-5}$ \\
Union2.....& $-0.66\pm0.03(1\sigma)\pm0.07(2\sigma)$	& $1.55\pm0.19(1\sigma)\pm0.38(2\sigma)$	& $0.74\pm0.17(1\sigma)\pm0.34(2\sigma)$	& 543.22 \\
Union2 + CMB/BAO.....& $-0.63\pm0.01(1\sigma)\pm0.03(2\sigma)$	& $1.36\pm0.02(1\sigma)\pm0.05(2\sigma)$	& $0.86\pm0.03(1\sigma)\pm0.10(2\sigma)$	& 543.98\\
\hline  
& & \quad \quad  \quad \quad{$\Omega_{k0} = 0$} & &\\
\hline 
CMB/BAO....& $-1.39\pm0.02(1\sigma)\pm0.04(2\sigma)$	& $2.55\pm0.03(1\sigma)\pm0.07(2\sigma)$	& $1.20\pm0.05(1\sigma)\pm0.10(2\sigma)$	& $1.51 \times 10^{-3}$ \\
Union2....& $-0.66\pm0.03(1\sigma)\pm0.07(2\sigma)$	& $1.54\pm0.19(1\sigma)\pm0.38(2\sigma)$	& $0.75\pm0.17(1\sigma)\pm0.35(2\sigma)$	& 543.22 \\
Union2 + CMB/BAO.....& $-0.63\pm0.01(1\sigma)\pm0.03(2\sigma)$	& $1.35\pm0.02(1\sigma)\pm0.05(2\sigma)$	& $0.87\pm0.04(1\sigma)\pm0.10(2\sigma)$	& 544.00 \\
\hline  
& & {$\Omega_{k0} = 0$ and $q(z>>1) = 1/2$} & &\\
\hline 
CMB/BAO.....& $-0.16\pm0.05(1\sigma)\pm0.10(2\sigma)$ & \quad \quad \quad \quad \quad \quad   -- &$0.32\pm0.10(1\sigma)\pm0.20(2\sigma)$ & 1.76 \\
Union2.....& $-0.58\pm0.04(1\sigma)\pm0.09(2\sigma)$&\quad \quad \quad \quad  \quad \quad   -- &$1.16\pm0.08(1\sigma)\pm0.18(2\sigma)$& 543.79 \\
Union2 + CMB/BAO.....& $-0.36\pm0.03(1\sigma)\pm0.06(2\sigma)$ & \quad \quad \quad \quad  \quad \quad  -- & $0.71\pm0.06(1\sigma)\pm0.12(2\sigma)$& 591.59 \\
\hline  \hline
\end{tabular} 
\end{center} 
\end{table*}


\emph{Conclusions}--We have performed a cosmographic analysis to estimate the epoch of cosmic acceleration without making any assumption about the underlying gravitational theory and energy components of the universe. We have used in our analyses the most recent SNe Ia data, the so-called Union2 sample of 557 events in the range $0.015 < z < 1.4$, along with estimates involving the product of the CMB acoustic scale $\ell_{A}$ and the BAO dilation scale ($D_V$) at two different redshifts ($z = 0.2$ and $z = 0.35$). 

We have shown that the complementarity of these two data sets (see Figs. 1a and 2b) poses tight constraints on $z_{acc}$ (with uncertainties of $\simeq 15\%$), with the current estimates being in good agreement with the standard $\Lambda$CDM scenario with $\Omega_m \sim 0.27$. In agreement with previous analyses, our study strongly favors a Universe with a recent acceleration ($q_0 < 0$) and an earlier decelerated phase ($q_1 > 0$). We have also checked the influence of curvature on the estimates of the deceleration parameter and accelerating redshift (see Table I).

Finally, it is worth emphasizing that the analyses performed here  clearly provide independent evidence for a dynamical model in which a deceleration/acceleration transition happened at $z \simeq 1$. From the combination of data used in our analyses it is also possible to detect a cosmic acceleration today with great confidence, even allowing for arbitrary curvature.


\acknowledgments

Th authors acknowledge CAPES and CNPq for the grants under which this work was carried out.

\end{document}